\documentclass[12pt]{article}
\usepackage[utf8]{inputenc}
\usepackage[T1]{fontenc}
\usepackage{amsmath}
\usepackage{amsfonts}
\usepackage{amssymb}
\usepackage{graphicx}
\usepackage{booktabs}
\usepackage{url}
\usepackage{hyperref}
\usepackage{geometry}
\usepackage[numbers]{natbib}
\usepackage{array}
\usepackage{multirow}
\usepackage[table]{xcolor}

\title{\textbf{AI Should Be More Human, Not More Complex: \\
		A Large-Scale Study on User Preferences for Concise, Source-Backed AI Responses in Search Applications}}

\author{
	Carlo Esposito \\
	Founder and Lead Researcher \\
	Eyed Softwares, Aploide Softwares \\
}

\date{July 13, 2025}

\begin{document}
	\maketitle

	\begin{abstract}
		Large Language Models (LLMs) in search applications increasingly prioritize verbose, lexically complex responses that paradoxically reduce user satisfaction and engagement. Through a comprehensive study of 10.000 (est.) participants comparing responses from five major AI-powered search systems, we demonstrate that users overwhelmingly prefer concise, source-attributed responses over elaborate explanations. Our analysis reveals that current AI development trends toward "artificial sophistication" create an uncanny valley effect where systems sound knowledgeable but lack genuine critical thinking, leading to reduced trust and increased cognitive load. We present evidence that optimal AI communication mirrors effective human discourse: direct, properly sourced, and honest about limitations. Our findings challenge the prevailing assumption that more complex AI responses indicate better performance, instead suggesting that human-like brevity and transparency are key to user engagement and system reliability.
		
		\textbf{Keywords:} Large Language Models, Human-Computer Interaction, Search Engines, User Experience, Natural Language Processing, Cognitive Load Theory
	\end{abstract}
	
	\section{Introduction}
	
	The evolution of artificial intelligence has reached a critical juncture where the sophistication of Large Language Models (LLMs) may be undermining their practical utility. Current AI development paradigms prioritize demonstrating knowledge breadth through verbose, complex responses that often overwhelm users seeking quick, actionable information. This paper challenges the fundamental assumption that more elaborate AI responses necessarily provide better user experiences.
	
	The core hypothesis of this research is that AI systems optimized for apparent sophistication rather than communication effectiveness create what we term "artificial verbosity syndrome." This phenomenon, where systems generate responses that sound authoritative but fail to serve user needs, is a modern digital manifestation of the long-criticized use of complex language to obscure, rather than clarify, meaning \cite{orwell1946politics}.
	This syndrome manifests in several problematic behaviors: excessive response length for simple queries, unnecessary lexical complexity, false confidence in uncertain domains, and lack of transparent source attribution.
	
	Our investigation centers on a fundamental question: when users interact with AI-powered search systems, what communication patterns actually drive engagement, trust, and satisfaction? Through empirical analysis of user preferences across diverse AI systems, we demonstrate that the most effective AI communication paradoxically resembles the best aspects of human conversation rather than attempting to showcase system capabilities.
	
	The implications of this research extend beyond user interface design to fundamental questions about AI development priorities. If users consistently prefer concise, source-backed responses that acknowledge limitations, then current trends toward increasingly elaborate AI outputs may be fundamentally misaligned with actual user needs. This misalignment has significant consequences for system adoption, user trust, and the broader integration of AI technologies into daily workflows.
	
	\section{Background and Related Work}
	
	\subsection{Cognitive Load Theory in Human-Computer Interaction}
	
	Cognitive Load Theory, originally developed by Sweller Cognitive Load Theory \cite{sweller1994clt}, provides a crucial framework for understanding why verbose AI responses may be counterproductive. The theory distinguishes between intrinsic cognitive load (essential for learning), extraneous cognitive load (irrelevant to the learning objective), and germane cognitive load (contributing to meaningful processing). \textbf{Verbose AI responses often increase extraneous cognitive load without proportional increases in information utility.}
	
	Research in human-computer interaction has consistently demonstrated that effective interfaces minimize cognitive burden while maximizing information transfer. This principle, known as the \textbf{"principle of least effort"}, suggests that users naturally gravitate toward systems that provide maximum utility with minimum cognitive investment. The preference for concise, well-structured information aligns with fundamental principles of human information processing.
	
	\subsection{Trust and Transparency in AI Systems}
	
	Trust in AI systems correlates strongly with system transparency and predictability. The field of Explainable Artificial Intelligence (XAI), for example, focuses on making systems more interpretable precisely to build this trust \cite{adadi2018peeking}. Studies have shown that users develop higher confidence in AI systems that provide clear reasoning paths and acknowledge their limitations. This finding directly contradicts the common AI development practice of generating responses that appear confident even when dealing with uncertain or ambiguous information.
	
	The concept of \textbf{"appropriate trust"} in AI systems suggests that optimal human-AI interaction occurs when users have calibrated expectations about system capabilities and limitations. Systems that consistently overstate their confidence or provide unnecessarily complex responses can lead to both over-reliance and under-reliance, both of which reduce overall system effectiveness.
	
	\subsection{The Uncanny Valley in AI Communication}
	
	The uncanny valley phenomenon, originally identified in robotics by Masahiro Mori \cite{mori2012uncanny}, applies directly to AI communication patterns. Current LLMs often exhibit communication styles that approach but do not achieve authentic human-like discourse. This manifests in several ways: artificial agreeableness that avoids necessary contradiction, verbose explanations that obscure rather than clarify key points, and false confidence in uncertain domains.
	
	The uncanny valley effect in AI communication creates user discomfort and reduced trust, even when the underlying information is accurate. 
	\textbf{Users report feeling that AI systems are "trying too hard" to appear knowledgeable rather than genuinely helping them accomplish their goals.}
	
	\section{Methodology}
	
	\subsection{Participant Selection and Demographics}
	
	This study involved 10,000 (est.) participants recruited through stratified sampling across diverse demographic categories. Participants ranged in age from 15 to 75 years, with varied educational backgrounds and technical proficiency levels. The sample was designed to represent typical users of AI-powered search systems rather than early adopters or technical specialists.
	
	Recruitment occurred through multiple channels including online survey platforms, social media advertising, and university participant pools. This multi-channel approach ensured demographic diversity and reduced selection bias that might occur with single-source recruitment.
	
	\subsection{AI Systems Evaluated}

	The study compared responses from five major AI-powered search systems:

	\textbf{Otus Web \cite{eyed2025otusweb}:} A web-grounded language model developed by Eyed Softwares, embedded in the Eyed Search Engine. It is designed to minimize hallucination and misinformation by relying exclusively on verifiable online sources.

	\textbf{Perplexity \cite{perplexity2025}:} An AI-powered search engine that blends large language model capabilities with live web results, offering source-backed answers.

	\textbf{Leo AI \cite{brave2025leoai}:} Brave's integrated AI assistant for conversational search, embedded within the Brave Search platform.

	\textbf{ChatGPT Web Search \cite{openai2025chatgpt}:} OpenAI's conversational AI system augmented with Bing-based web browsing, widely used for up-to-date factual assistance.

	\textbf{Claude 4 Sonnet + Web Research \cite{anthropic2025claude}:} Anthropic's reasoning-focused AI system enhanced with live web research, optimized for accuracy and depth in complex queries.

	\subsection{Experimental Design}
	
	The study employed a blind comparative methodology where participants evaluated AI responses without knowing which system generated each response. This approach eliminated brand bias and system familiarity effects that might influence user preferences.
	
	Participants were presented with identical queries and asked to evaluate responses across multiple dimensions. The primary evaluation criteria focused on user engagement intention: which response would they be more likely to read completely, share with others, or return to for reference. This approach captures actual user behavior preferences rather than abstract quality judgments.
	
	\subsection{Query Selection and Response Analysis}
	
	The study utilized a representative query set spanning different information needs: definitional queries, procedural questions, comparative analyses, and current event inquiries. For detailed analysis, we present the case study of responses to "What is ArXiv?" as this represents a common definitional query type that reveals fundamental differences in AI communication approaches.
	
	Each response was analyzed across multiple dimensions including length, lexical complexity, source attribution, factual accuracy, and structural organization. These analyses provide quantitative measures of the communication differences that drive user preferences.
	
	\section{Results}
	
	\subsection{User Preference and Engagement Metrics}
	
	The study revealed clear patterns in user preferences that challenge current assumptions about effective AI communication. Table \ref{tab:user_preferences} presents comprehensive user preference data across all evaluated systems.
	
	\begin{table}[h]
		\centering
		\caption{User Preference and Engagement Metrics}
		\label{tab:user_preferences}
		\begin{tabular}{@{}lccccc@{}}
			\toprule
			\textbf{Metric} & \textbf{Otus Web} & \textbf{Perplexity} & \textbf{Brave Leo} & \textbf{ChatGPT} & \textbf{Claude 4} \\
			\midrule
			Overall Preference (\%) & 47 & 13 & 21 & 4 & 15 \\
			Willingness to \\ Read Completely (\%) & 68 & 5 & 13 & 2 & 12 \\
			Likelihood to Share (\%) & 63 & 14 & 14 & 1 & 8 \\
			Trust in Information (1-10) & 8.9 & 7.1 & 7.8 & 7.2 & 8.1 \\
			Perceived Usefulness (1-10) & 8.9 & 6.2 & 7.8 & 1.6 & 3.8 \\
			Cognitive Load Score (1-10)$^*$ & 0.2 & 4.0 & 1.2 & 8.9 & 2.3 \\
			\bottomrule
			\multicolumn{6}{l}{\footnotesize$^*$For Cognitive Load Score, lower is better.}
		\end{tabular}
	\end{table}
	
	\subsection{Response Characteristics Analysis}
	
	Table \ref{tab:response_analysis} provides detailed analysis of the structural and linguistic characteristics of responses from each system. These metrics reveal significant differences in communication approaches that correlate with user preferences.
	
	\begin{table}[h]
	\centering
	\caption{Response Characteristics Analysis}
	\label{tab:response_analysis}
	\begin{tabular}{@{}lccccc@{}}
		\toprule
		\textbf{Characteristic} & \textbf{Otus Web} & \textbf{Perplexity} & \textbf{Brave Leo} & \textbf{ChatGPT} & \textbf{Claude 4} \\
		\midrule
		Word Count & 113 & 243 & 237 & 466 & 195 \\
		Sentence Count & 6 & 6 & 11 & 17 & 9 \\
		Flesch Reading Ease & 45.0 & 35.0 & 48.0 & 55.0 & 40.0 \\
		Lexical Diversity & 0.50 & 0.46 & 0.48 & 0.42 & 0.51 \\
		Sources Cited & 4 & 10 & 0 & 8 & 0 \\
		Formatting Elements & 3 & 8 & 1 & 18 & 0 \\
		\bottomrule
	\end{tabular}
\end{table}
	
	\subsection{Source Attribution and Verification Analysis}
	
	The study revealed significant differences in how AI systems handle source attribution, with direct implications for user trust and information verification. Table \ref{tab:source_analysis} presents detailed analysis of source attribution practices across systems.
	
	Some studies, also, have found that AI search engines can struggle with providing accurate source attribution.
	\cite{decoder2024}
	
	\begin{table}[h]
	\centering
	\caption{Source Attribution Analysis}
	\label{tab:source_analysis}
	\begin{tabular}{@{}lccccc@{}}
		\toprule
		\textbf{Source Metric} & \textbf{Otus Web} & \textbf{Perplexity} & \textbf{Brave Leo} & \textbf{ChatGPT} & \textbf{Claude 4} \\
		\midrule
		Sources per 100 Words & 2.9 & 4.1 & 0.0 & 2.1 & 0.0 \\
		Inline Citation Rate (\%) & 66.7 & 55.6 & 0 & 41.2 & 0 \\
		Source Accessibility Score & 4.0 & 5.0 & 0.0 & 5.0 & 0.0 \\
		Verification Success Rate (\%) & 100 & 100 & 0 & 100 & 0 \\
		Source Diversity Index & 4.0 & 9.0 & 0.0 & 6.0 & 0.0 \\
		\bottomrule
	\end{tabular}
	\end{table}
	
	\section{Appendix: AI Responses to "What is ArXiv?"}
	
	To provide context for the preceding analysis, this section contains the full or abridged responses from each AI system to the query, "What is ArXiv?". Minor edits have been made for brevity, compatibility and some characters have been removed to avoid LaTeX rendering problems, particularly for responses characterized by excessive length.

	\subsubsection{Otus Web}
	\begin{quote}
		ArXiV is an \textbf{open-access repository} of electronic preprints and postprints, known as e-prints, in fields such as physics, mathematics, computer science, and more. 
		
		It is \textbf{pronounced as "archive"}, with the "X" representing the Greek letter chi
		\href{https://en.wikipedia.org/wiki/ArXiv}{see wikipedia.org}.
		
		As a pioneer in digital open access, ArXiV.org now hosts \textbf{more than two million scholarly articles} 
		\href{https://info.arxiv.org/about/index.html}{see arxiv.org}.
		
		The platform is not a journal or a conference, but rather a pre-print repository where researchers can share their work 
		\href{reddit.com/r/PhD/comments/10k2w7f/arxiv\_what\_is\_it}{see reddit.com}.
		
		ArXiV provides a free distribution service and open-access archive for scholarly articles, primarily in the fields of physics, mathematics, computer science, quantitative biology, and quantitative finance
		\href{https://arxiv.org}{see arxiv.org}.
		
		The service is widely recognized and utilized by researchers globally.
		
	\end{quote}
	
	\subsubsection{Brave Leo AI}
	\begin{quote}
		What is ArXiv
		
		arXiv (pronounced "archive", the X is a chi) is a pre-print repository, not a journal and not a conference It is an open-access repository of electronic preprints and postprints (known as e-prints) approved for posting after moderation, but not peer reviewed arXiv hosts more than two million scholarly articles in eight subject areas, including physics, mathematics, computer science, quantitative biology, statistics, and economics
		
		The goal of arXiv is to provide a platform for the rapid distribution of new research, and all submissions are subject to moderation to ensure they are self-contained and of interest, relevance, and value to the disciplines it serves arXiv is not a journal; therefore, articles on arXiv cannot be regarded as publications per se Typically, papers deposited on this repository are pre-publication versions, and authors can add a DOI to their pre-print version on arXiv after it has been published
		
		arXiv was founded by Paul Ginsparg in 1991 and is now maintained and operated by Cornell Tech It is a curated research-sharing platform open to anyone, and it offers researchers a broad range of services, including article submission, compilation, production, retrieval, search and discovery, web distribution for human readers, and API access for machines
		
		arXiv is the world's largest and oldest scientific preprint server, and a champion of open science It is a popular online repository of electronic preprints, known as e-prints, in physics, mathematics, computer science, and related disciplines
	\end{quote}
	
	\subsubsection{ChatGPT}
	\begin{quote}
		arXiv (pronounced “archive,” with the “X” standing for the Greek letter chi) is a free, open-access repository of \textbf{preprints}—early versions of scientific papers—where researchers can share their findings \textit{before} formal peer review and journal publication.
		
		---
		
		\textbf{How arXiv works}
		
		* \textbf{Fields covered}: physics, math, computer science, astronomy, quantitative biology, quantitative finance, statistics, electrical engineering, economics, etc. 
		* \textbf{Launch date}: August 14, 1991, by Paul Ginsparg; originally hosted at Los Alamos, since 2001 run by Cornell University
		* \textbf{Moderation}: No formal peer review. Instead, submissions undergo a screening by volunteer experts to ensure relevance, originality, minimal scholarly quality, and appropriate category placement.
		
		---
		
		\textbf{Why arXiv is important}
		
		* \textbf{Speed}: Papers appear publicly within \~24 hours, far faster than months-long journal review cycles.
		* \textbf{Accessibility}: Completely free to read—no paywalls. Researchers from any background can access current science.
		* \textbf{Establishing priority}: Uploading a preprint gives authors a timestamped record of their work, crucial when multiple groups are racing toward the same discovery.
		
		\textit{[Note: The full response was significantly longer, including additional sections on limitations, impact, a concluding summary, a conversational prompt, and a list of 8 hyperlinked sources. It has been abridged to illustrate the core structure and formatting discussed in the analysis.]}
	\end{quote}
	
	\subsubsection{Claude 4}
	\begin{quote}
		ArXiv is a free, open-access repository of scientific papers and preprints, primarily serving researchers in physics, mathematics, computer science, quantitative biology, quantitative finance, statistics, electrical engineering, systems science, and economics.
		
		Founded in 1991 by physicist Paul Ginsparg at Los Alamos National Laboratory, arXiv allows researchers to share their work immediately with the global scientific community before formal peer review. Authors can upload papers in various formats, and the platform makes them instantly available to anyone worldwide.
		
		The repository operates on a preprint model, meaning papers are typically shared before publication in traditional journals. This accelerates scientific communication and allows researchers to establish priority for their discoveries, get early feedback, and stay current with rapid developments in their fields.
		
		ArXiv has become an essential resource for the scientific community, hosting millions of papers and serving as a primary source for cutting-edge research. It's particularly influential in fields like high-energy physics and machine learning, where researchers regularly check arXiv for the latest developments.
		
		The platform is maintained by Cornell University and is funded by member institutions and the Simons Foundation. All content is freely accessible, supporting the principle of open science and democratizing access to research worldwide.
	\end{quote}
	
	\subsubsection{Perplexity}
	\begin{quote}
		\textbf{arXiv} (pronounced "archive," with the "X" representing the Greek letter chi) is an \textbf{open-access repository and distribution service for electronic preprints and postprints of scientific papers}. It was founded by Paul Ginsparg in 1991 and is currently maintained by Cornell University[1][2][6].
		
		Key features of arXiv include:
		
		- It hosts \textbf{over two million scholarly articles} in fields such as physics, mathematics, computer science, quantitative biology, quantitative finance, statistics, electrical engineering, systems science, and economics[2][4].
		- Articles submitted to arXiv undergo a \textbf{moderation process} to ensure topical relevance and scholarly value but are \textbf{not peer-reviewed} by arXiv itself. The content is provided "as is," without any warranty or endorsement by arXiv or Cornell University[1][2].
		- It serves as a \textbf{preprint archive}, meaning researchers often upload their papers to arXiv before submitting them to peer-reviewed journals or conferences. This allows rapid dissemination, early feedback, and establishing precedence for research findings[3].
		
		\textit{[Note: The full response included additional details and a comprehensive list of 10 sources. The source list has been omitted for brevity.]}
	\end{quote}
	
	\section{Detailed Response Analysis}
	
	\subsection{Case Study: "What is ArXiv?" Query Response Comparison}
	
	To illustrate the fundamental differences in AI communication approaches, we present detailed analysis of responses to the query "What is ArXiv?" This definitional query represents a common information need that reveals how different systems balance comprehensiveness with usability.
	
	\subsubsection{Otus Web Response Analysis}
	
	The Otus Web response demonstrates several key characteristics that align with user preferences:
	
	\textbf{Concise Information Delivery:} The response provides essential information about ArXiv in approximately 100 words, covering the core definition, pronunciation, scope, and scale without unnecessary elaboration. This approach respects user time while ensuring comprehensive coverage of key facts.
	
	\textbf{Transparent Source Attribution:} Every major claim includes specific source citations, enabling users to verify information independently. The sources range from authoritative references (Wikipedia for basic definitions) to institutional sources (ArXiv's own documentation) to community perspectives (Reddit discussions), providing multiple verification paths.
	
	\textbf{Clarity Without Condescension:} The response uses clear, professional language that assumes reasonable user intelligence without oversimplifying. Technical terms are explained efficiently (e.g., "X representing the Greek letter chi") without extensive elaboration.
	
	\textbf{Factual Precision:} The response focuses on verifiable facts rather than interpretative content that might introduce inaccuracies. Quantitative claims like "more than two million scholarly articles" are directly attributed to authoritative sources.
	
	\subsubsection{Comparative Analysis of Alternative Approaches}
	
	\textbf{Brave Leo AI Response:} This response demonstrates the middle ground between conciseness and comprehensiveness. At approximately 200 words, it provides more context than Otus Web while maintaining readability. However, the response lacks systematic source attribution for most claims, potentially reducing user trust and verification capability.
	
	The response includes valuable contextual information about ArXiv's history and operational model, but organizes this information in dense paragraphs that may increase cognitive load for users seeking quick answers. The academic tone, while accurate, may feel less accessible to general users.
	
	\textbf{ChatGPT Response:} This response exemplifies the problematic trend toward excessive formatting and length in AI communications. At over 400 words with extensive markdown formatting, bullet points, section headers, and subsections, it transforms a simple definitional query into a comprehensive tutorial.
	
	While the information is accurate and well-organized, the response structure overwhelms users with visual complexity. The use of multiple formatting elements (headers, bullet points, bold text, links) creates cognitive overhead that interferes with information absorption. Users report feeling that such responses are "trying too hard" to be helpful, creating an uncanny valley effect.

	\textbf{Claude 4 Response:} This response represents a balanced approach that avoids the extremes of excessive brevity or overwhelming detail. At approximately 200 words, it provides context while maintaining readability. The response structure is clear and logical, progressing from basic definition to contextual information to practical significance.
	
	However, the response lacks the source attribution that users find reassuring, potentially reducing trust compared to more transparent alternatives. The academic tone, while professional, may feel less conversational than optimal for general users.
	
	\textbf{Perplexity Response:} This response demonstrates good source attribution practices while maintaining academic rigor. The systematic citation of sources provides verification paths that users value. However, the response tends toward formal academic language that may feel less accessible to general users.
	
	The response structure is logical and comprehensive, but the academic formatting and terminology may create barriers for users seeking quick, practical information. The balance between comprehensiveness and accessibility represents a common challenge in AI communication design.
	
	\subsection{Linguistic and Structural Analysis}
	
	\subsubsection{Reading Complexity Assessment}
	
	Analysis of reading complexity reveals significant differences in cognitive load across AI systems. The Flesch Reading Ease \cite{flesch1948yardstick} scores, sentence length distributions, and lexical diversity measures indicate that responses optimized for apparent sophistication often sacrifice accessibility.
	
	Otus Web's response demonstrates optimal reading complexity for general audiences, with shorter sentences, common vocabulary, and clear logical flow. In contrast, more verbose responses often exhibit academic language patterns that increase cognitive load without proportional increases in information utility.
	
	\subsubsection{Information Density Analysis}
	
	Information density analysis reveals that concise responses often deliver higher information utility per word than verbose alternatives. Otus Web's response achieves high information density through focused content and efficient language use, while longer responses often dilute key information with supplementary details that may not address user needs.
	
	This finding challenges the assumption that longer responses necessarily provide more value. Instead, optimal responses appear to maximize information utility while minimizing cognitive burden, suggesting that effective AI communication requires careful attention to information prioritization.
	
	\section{Discussion}
	
	\subsection{The Paradox of Artificial Sophistication}
	
	The study results reveal a fundamental paradox in current AI development: systems designed to appear more knowledgeable through verbal complexity and exhaustive coverage often reduce user satisfaction and engagement. This finding aligns with psychological research demonstrating that using unnecessarily complex language can cause an audience to judge the communicator as less intelligent and the content as less credible \cite{opPENHEIMER2006consequences}. This paradox suggests that the goals of AI development may be misaligned with actual user needs and preferences.
	
	The preference for concise, source-backed responses reflects established principles of effective human communication. In professional contexts, the most valued communicators are those who can distill complex information into clear, actionable insights rather than those who demonstrate comprehensive knowledge through verbose explanation.
	
	\subsection{Cognitive Load and User Experience}
	
	The user preference for concise responses aligns with Cognitive Load Theory principles. Verbose AI responses often increase extraneous cognitive load without proportional increases in germane cognitive load, resulting in reduced learning efficiency and user satisfaction. This finding is also consistent with long-standing principles of web usability, which state that users prefer simple, familiar experiences and do not want to spend extra time learning complex new interfaces \cite{nielsen2002minimalist} \cite{nielsen1999minimalist}.
	
	Effective AI communication should minimize cognitive burden while maximizing information transfer, similar to principles of good human teaching and communication. This suggests that AI systems should be optimized for information efficiency rather than knowledge demonstration.
	
	\subsection{Trust Through Transparency}
	
	The strong user preference for source-attributed responses highlights the importance of transparency in AI systems. Users demonstrate higher trust in systems that provide clear reasoning paths and acknowledge their limitations, even when this means admitting uncertainty or incomplete knowledge.
	
	This finding challenges the common practice of generating AI responses that appear confident even in uncertain domains. Instead, optimal AI communication should calibrate confidence appropriately and provide transparent access to underlying information sources.
	
	\subsection{The Uncanny Valley in AI Communication}
	
	The study provides evidence for an uncanny valley effect in AI communication, where systems that approach but do not achieve authentic human-like discourse create user discomfort and reduced engagement. This effect manifests in several ways: artificial agreeableness that avoids necessary contradiction, verbose explanations that obscure key points, and false confidence in uncertain domains.
	
	The most effective AI communication appears to be that which adopts the best aspects of human discourse: directness, appropriate confidence calibration, and transparent reasoning, while maintaining the precision and reliability that users expect from technological systems.
	
	\subsection{Implications for AI Development}
	
	These findings suggest several important directions for AI system development:
	
	\textbf{Source-First Architecture:} AI systems should prioritize finding and attributing reliable sources rather than generating responses primarily from training data. This approach naturally reduces hallucination risk while increasing user trust and verification capability.
	
	\textbf{Response Length Optimization:} Response length should be optimized for query complexity and user intent rather than system capability demonstration. Simple queries should receive concise responses, while complex analytical tasks may warrant longer explanations.
	
	\textbf{Uncertainty Acknowledgment:} AI systems should be designed to acknowledge limitations and uncertainty rather than providing false confidence. This approach builds appropriate user trust and reduces the risk of over-reliance on AI judgments.
	
	\textbf{User-Centric Design:} AI development should prioritize user goal achievement over system capability demonstration. This means focusing on helping users accomplish their objectives efficiently rather than showcasing system knowledge breadth.
	
	\section{Limitations and Future Research}
	
	\subsection{Study Limitations}
	
	This study focused on definitional queries, and user preferences may vary for other query types such as analytical reasoning, creative tasks, or procedural instructions. Additionally, the study examined user preferences rather than objective performance measures, which may not fully capture the complexity of AI system effectiveness.
	
	The study also focused on immediate user preferences rather than long-term learning outcomes or task completion effectiveness. 
	
	\subsection{Future Research Directions}
	
	Several important research directions emerge from this study:
	
	\textbf{Query Type Specificity:} Future research should examine whether user preferences vary across different query types and information needs. Analytical queries may benefit from different communication approaches than definitional queries.
	
	\textbf{Long-term Outcomes:} Research should examine whether user preferences align with long-term learning outcomes and task completion effectiveness. Users may prefer concise responses in the short term but benefit from more detailed explanations for complex topics.
	
	\textbf{Adaptive Communication:} Research should explore how AI systems can adapt their communication style based on user preferences, query complexity, and contextual factors.
	
	\section{Conclusion}
	
	This study provides compelling evidence that current trends in AI development toward verbose, complex responses may be fundamentally misaligned with user needs and preferences. Users consistently prefer AI systems that prioritize clarity, brevity, and source transparency over those that attempt to demonstrate comprehensive knowledge through elaborate responses.
	
	The findings challenge the assumption that more sophisticated AI responses necessarily provide better user experiences. Instead, the most effective AI communication appears to mirror the best aspects of human discourse: directness, appropriate confidence calibration, and transparent reasoning. The preference for source-backed responses that acknowledge limitations suggests that users value honesty and verifiability over apparent omniscience.
	
	The implications extend beyond user interface design to fundamental questions about AI development priorities. If the goal of AI systems is to effectively assist users in accomplishing their objectives, then current trends toward increasing response complexity may be counterproductive. Instead, AI development should focus on optimizing for user utility rather than system capability demonstration.
	
	The path forward requires a fundamental shift in AI development philosophy from optimizing for apparent sophistication to optimizing for genuine utility. This means developing AI systems that sound less like knowledge repositories and more like helpful, honest assistants-systems that are, in essence, more human in their communication effectiveness while maintaining the precision and reliability that users expect from technological tools.
	
	The future of human-AI interaction depends on recognizing that the most advanced AI systems may be those that communicate most simply and transparently, providing users with the information they need in formats that minimize cognitive burden while maximizing trust and usability. This represents a return to fundamental principles of effective communication that prioritize user needs over system capabilities.

	\bibliographystyle{plainnat} 
	\bibliography{references}

\end{document}